\documentclass[12pt]{iopart}

\usepackage{graphicx}
\usepackage{color}
\begin{document}

\title[Universal fluctuations of generalized diffusivity on percolation clusters]{Universal fluctuations  and ergodicity  of generalized diffusivity on critical percolation clusters}

\author{Adrian Pacheco-Pozo$^{1}$ and Igor M. Sokolov$^{1,2}$}

\address{$^1$ Institut f{\"u}r Physik, Humboldt-Universit{\"a}t zu Berlin, Newtonstra{\ss}e 15, D-12489 Berlin, Germany}
\address{$^2$ IRIS Adlershof, Zum Großen Windkanal 2, D-12489 Berlin, Germany}
\ead{adrian.pacheco@physik.hu-berlin.de and igor.sokolov@physik.hu-berlin.de}
\vspace{10pt}
\begin{indented}
\item[]
\end{indented}

\begin{abstract}
Despite a long history and a clear overall understanding of properties of random walks on an incipient infinite cluster in percolation, some important information on it seems to be missing in the literature. In the present work, we revisit the problem by performing massive numerical simulations for (sub)diffusion of particles on such clusters. Thus, we discuss the shape of the probability density function (PDF) of particles' displacements, and the way it converges to its long-time limiting scaling form. Moreover, we discuss the properties of the mean squared displacement (MSD) of a particle diffusing on the infinite cluster at criticality. This one is known not to be self-averaging. We show that the fluctuations of the MSD in different realizations of the cluster are universal, and discuss the properties of the distribution of these fluctuations. These strong fluctuations coexist with the ergodicity of subdiffusive behavior in the time domain. The dependence of the relative strength of fluctuations in time-averaged MSD on the total trajectory length (total simulation time) is characteristic for diffusion in a percolation system and can be used as an additional test to distinguish this process with disorder-induced memory from processes with otherwise similar behavior, like fractional Brownian motion with the same value of the Hurst exponent. 
\end{abstract}

%
%
%
%
%

\section{Introduction}

Percolation models (see e.g. \cite{Stauffer1994}) constitute the simplest class of models of structurally disordered systems, and have a host of applications where they can be considered as relatively faithful representations that could, up to a certain degree, reproduce the properties of realistic physical systems, see e.g. \cite{Saberi2015,Araujo2014} for a review. The study of percolation problems started with the seminal work of Broadbent and Hammersley \cite{Broadbent1957}. Broadbent and Hammersley have stated a clear difference between percolation and diffusion processes, with concentration at the first. The application which stimulated the whole line of investigation was fluid's permeation through porous media \cite{Araujo2014}, see also the account of G. R. Grimmett available online \cite{Grimmett} (see also \cite{Grimmett1999}), and immediate further applications were to electric conductivity of strongly disordered media, see \cite{Kirkpatrick1973} for an early review. Although the ideas of percolation theory rapidly penetrated physics (e.g. motivated by P. W. Anderson's approach to a quantum localization problem \cite{Anderson1958}), the true interest to diffusion in percolating systems only came considerably later, probably with the work of Brandt \cite{Brandt1975}, and sparkled after it was  popularized by P.-G. de Gennes \cite{deGennes} by  introducing the term ``ant in a labyrinth'' to refer to this peculiar type of diffusion. Being interesting on its own rights as a mathematical model, diffusion on percolation clusters has multiple application in many branches of natural science including biology (see \cite{Havlin2002,Saxton2010,Baumann2010}). Due to the multiplicity of applications, percolation models now come in a host of different ramifications and flavors \cite{Saberi2015}. In what follows we concentrate on the simplest and oldest variants of Bernoulli percolation on a lattice. The results shown in our work are pertinent to bond percolation on a square lattice, where the diffusion is modeled by a continuous-time random walk corresponding to a ``myopic ant'' according to de Gennes classification, but were checked to be independent both on the type of the problem (bond or site percolation) and of the ant (myopic or blind), as they should be due to universality. 

Below the percolation threshold, only finite clusters of connected sites are present in the system, the motion of the walker is bounded to the cluster where it started, and the mean squared displacement (MSD) stagnated at longer times. Above the percolation threshold, an infinite cluster (which is unique under the conditions investigated) appears. The finite clusters, which are present along with the infinite one,  may or may not be of interest depending on the particular situation \cite{Mardoukhi2015,Mardoukhi2018}. An example for the second situation is the case of a porous system, where the infinite cluster of voids is filled by a fluid medium (and therefore can be the only part of the system where the diffusion is possible), while finite ones are unavailable for diffusion.  In what follows we concentrate only on this situation, when diffusion takes place on the infinite cluster, leaving finite clusters out of attention. 

Above the critical concentration, the diffusion on the infinite percolation cluster (which we will simply call ``percolation cluster'' in what follows) is a typical example of a diffusion process showing homogenization. This means that at long times or large scales the properties of such diffusion are described by those of some effective normal diffusion process (mathematically equivalent to a Brownian motion). This is true both if the corresponding properties are considered under averaging over the ensemble of the realization of the percolation environment (this is what mathematicians call ``annealed invariance principle'') or in a single realization (``quenched invariance principle'') \cite{Barlow2016}. This means, that averaging over different realizations of disorder is essentially redundant. Physicists speak about this effect as ``self-averaging'' of the corresponding property. 

From the physical point of view, the self-averaging property, say, of the mean squared displacement (MSD) $\langle \mathbf{x}^2(t) \rangle$ (and of the corresponding effective diffusion coefficient $D^* = \langle \mathbf{x}^2(t) \rangle/2 d t$ for $t$ large with $d$ being the dimension of space) is of no wonder, since there exists a characteristic size of the inhomogeneities (``holes'' in the cluster), the correlation length $\xi$, above which the cluster can be considered as homogeneous on the average.

The structure of the infinite cluster below $\xi$ is known to be fractal, and diffusion on fractal systems often shows anomalous behavior at times considerably exceeding the time of a single step: 
\begin{equation}
 \langle \mathbf{x}^2(t) \rangle \propto t^{2/d_w},
 \label{eq:MSDFrac}
\end{equation}
with $d_w$ being the walk dimension (being different from its value $d_w=2$ for normal diffusion). For two-dimensional percolation systems discussed below the accepted value of $d_w$ is $d_w \approx 2.871$, and the behavior is subdiffusive. 

In the case where $p > p_c$, the MSD in the whole time domain behaves as \cite{Bunde2005}
\begin{equation}
\langle \mathbf{x}^2(t) \rangle \propto
    \left\{ 
            \begin{array}{lcc}
             t^{2/d_w} &   \mbox{for}  & t \ll t_{\xi} \\
             t &  \mbox{for} & t \gg t_{\xi},
             \end{array}
   \right.
\label{eq:MSD}
\end{equation}
where $t_{\xi} \sim \xi^{d_w}$ is the crossover time connected with the correlation length $\xi$. Together with the lattice spacing $a$, being the lowest space scale, the correlation length $\xi$ bounds the domain of length scales in which the homogenization does not take place. The behavior corresponding to length scales in this domain is inherited from the critical state, and is different from the one in the homogenization domain. For $p \to p_c^+$ the correlation length grows, and with this also the domain of the applicability of Eq.~(\ref{eq:MSDFrac}). The average in Eqs.~(\ref{eq:MSDFrac}) and (\ref{eq:MSD}) has then to be understood as a double average over thermal histories, i.e. trajectories of random walks in a given random environment $\langle ... \rangle_T$, and over the ensemble of such environments, i.e. incipient percolation clusters $C$, $\langle ... \rangle_C$: $\langle \mathbf{x}^2(t) \rangle =  \langle \langle \mathbf{x}^2(t)  \rangle_T\rangle_C$. Eq.~(\ref{eq:MSDFrac}) allows to define the effective coefficient $K^*$ of anomalous diffusion $K^* = \lim_{t \to \infty}  \langle \mathbf{x}^2(t) \rangle /t^{2/d_w}$, and to characterize the corrections to scaling by introducing a time-dependent diffusion coefficient $K(t) =  \langle \mathbf{x}^2(t) \rangle / t^{2/d_w}$ which tends to $K^*$ as $t$ grows. 

In the time of interest to fractal systems (80th - 90th years of the previous century) the incipient percolation cluster (i.e. the infinite cluster for concentrations above but asymptotically close to $p_c$) and diffusion thereon were mostly considered as ``just an example'' of typical fractal behavior, without paying too much attention to the peculiarity of the strong disorder present in the percolation case, with only a few exceptions.

While the existence of the quenched invariance principle suggests that for $t \gg t_{\xi}$ it doesn't matter whether a single realization of disorder, or many such realizations are considered, for $t \ll t_{\xi}$ this does not have to be the case. In particular, for the case of the diffusion on the incipient percolation cluster discussed above, one has $\xi \to \infty$ and thus $t_{\xi} \to \infty$. Hence, there is only a lower cutoff scale (lattice constant) and no upper characteristic scale ensuring homogenization. Then, for $t \to \infty$, i.e. for large typical displacements, the behavior of $K_T(t) = \langle \mathbf{x}^2(t) \rangle_T / t^{2/d_w}$ may follow one of the two alternative scenarios:
\begin{itemize}
 \item The value of $K_T(t) = \langle \mathbf{x}^2(t) \rangle_T / t^{2/d_w}$ tends to a deterministic limit, manifesting the existence of the correspondingly modified quenched invariance principle (convergence in probability).
 
 \item There is no convergence in probability. Then, due to the self-similarity (i.e. the absence of the characteristic scale) of the cluster, one could anticipate, that the \textit{distribution} of $K_T(t)$ gets to be independent of $t$, because, at long $t$, it is dominated by relatively large excursions from the origin, at which length scales where the lower cutoff length (lattice constant) ceases to play a role. Therefore a plausible alternative to convergence in probability is the convergence in distribution. 
\end{itemize}
One often speaks about the first situation as the case when the disorder is irrelevant, and about the second as the situation where it is relevant \cite{Aharony1996}. In critical percolation the disorder is relevant, and the distribution of displacements does not show self-averaging \cite{Bunde1995}, however many properties of the distribution of effective diffusivities were hardly reported on. 

We encountered this problem trying to make semi-quantitative estimates for the properties of the probability density function (PDF) of displacements for particle diffusing on a supercritical percolation landscapes in our work \cite{Pacheco2021}. To do so, we needed to know at least rough properties of the distribution $p(K_T)$, and were not able to find any hints on them in the available literature (although some related distributions, e.g. the one of resistance distance \cite{Harris1987}, were mentioned). The existence of considerable gaps in our understanding of the properties of diffusion on percolation clusters at criticality was quite astonishing given the very long history of the problem, and was our first motivation for the present work. Working on the properties of $p(K_T)$, we have found several other interesting properties of diffusion on percolation clusters. Thus, the MSD on a single realization of a percolation cluster essentially never ``forgets'' the initial condition. This property, however, coexists with the ergodicity of the MSD, under which the double (trajectory and cluster) average coincides with the time-averaged MSD over a single trajectory \cite{Meroz2013}. The convergence of the two averages to each other follows, however, a different pattern than, say, in the fractional Browinan motion (FBM) showing a very similar pattern of subdiffusive behavior. This difference is immediately connected with the presence of strong disorder, and can be used as an additional test to distinguish between the disorder-induced memory, and the intrinsic one (say, induced by the presence of slow modes). 

In what follows we revisit the problem of diffusion on a critical percolation cluster and report on the results of simulations concerning the probability density function (PDF) of displacements (averaged over the realizations of the clusters), on its convergence to a scaling form, on the distribution of the effective diffusion coefficient, and on the convergence of time-averaged mean squared displacements (TAMSD) and MSD under cluster averaging, by this filling at least some of the gaps mentioned.  

\section{The probability density function}

\begin{figure}[h!]
\centering
\def\svgwidth{0.65\columnwidth}
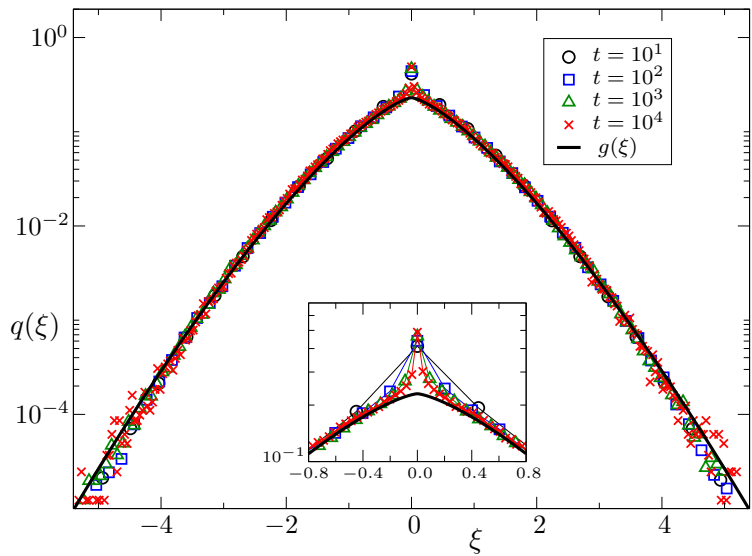
\caption{Time evolution of the PDF of the rescaled walker's displacement ($\xi = x / t^{ 1/ d_w}$) on a bond percolation landscape at criticality ($p = p_c = 0.5$), see text for details. 
The inset shows a close up of the central part of the PDF, exhibiting the same kind of convergence by narrowing of the central peak (``chupchik'') as found in strongly disordered systems showing homogenization, see \cite{Pacheco2021}.}
\label{fig:PDF}
\end{figure}

The PDF of the particles' displacements in bond percolation on a two-dimensional square lattice is shown in Fig.~\ref{fig:PDF}. The myopic-ant particle performs a continuous-time random walk on the cluster with equal transition rates to each of its available neighbors, see \ref{AppTraj} for details. These transition rates are put to unity thus defining the time scale of the problem.

Let us first discuss the representation of the PDF in Fig.~\ref{fig:PDF}, which is similar to the one used in  \cite{Pacheco2021}, and differs from the one adopted in early works like \cite{Havlin1985}. The PDF $p(x,y)$ (i.e. the correspondingly normalized probabilities to find a walker at the corresponding sites of the lattice, averaged over many realizations of the clusters) is an isotropic function of the coordinates, and is typically plotted as a function of $r=\sqrt{x^2 + y^2}$. The choice of plotting $2\pi r p(r)$ fully obscures the part close to the mode of the distribution, which is however distinctive for cases of strong disorder \cite{Pacheco2021}. In Fig.~\ref{fig:PDF} we show the plot of the ``cut'' of the PDF by a line passing through the origin, $p(x,0)$, from which the full PDF $p(x,y)$ can be easily restored assuming the isotropy. Plotted is the value of $q(\xi) = t^{2 /d_w} p(x,y=0)$ as a function of $\xi = x / t^{1 / d_w}$. The corresponding PDF shows clear scaling at its wings, and the convergence to the scaling form follows by narrowing of the central peak. This type of convergence was also found typical for systems with strong disorder showing homogenization to normal diffusion \cite{Pacheco2021}. 

The scaling part of the distribution is well described by a stretched exponential of the form
\begin{equation}
g(\xi) = a \exp ( - |\xi|^b )
\label{eq:scaled}
\end{equation}
with positive fitting parameters $a$ and $b$. Non-linear fitting of Eq.~(\ref{eq:scaled}) to the simulated PDF for the longest time $t=10^4$ performed by employing a Levenberg–Marquardt algorithm \cite{SciPy2020} gives the values for these parameters $a = 0.23$ and $b = 1.37$, and shows an extremely good agreement of the fit $g(\xi)$ (shown with a full line in Fig.~\ref{fig:PDF}) with the PDFs obtained from simulation. Note that the parameter $b$ of the stretched exponential is considerably smaller that the one found in \cite{Havlin1985}. We stress here that our PDF is a mean PDF obtained by averaging over the ensemble of clusters.

\section{Some properties of the MSD at criticality}

\begin{figure}[h!]
\centering
\def\svgwidth{0.65\columnwidth}
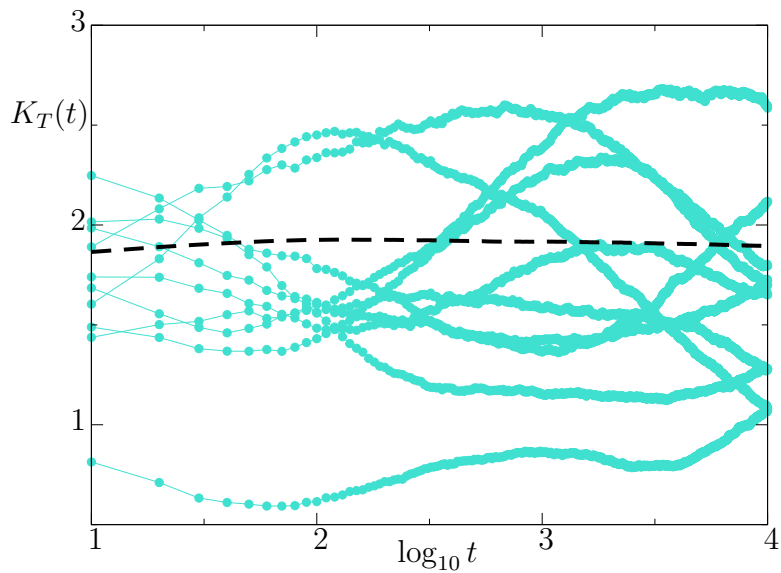
\caption{A comparison between the generalized diffusion coefficient $K_T(t)$ of ten realizations of the infinite percolation cluster (symbols) and the one obtained by landscape average over 1000 realizations of such clusters (dashed line). }
\label{fig:K_clusters}
\end{figure}

The behavior of $\langle \mathbf{x}^2(t) \rangle_T$ for single realizations of clusters is shown in Fig.~\ref{fig:K_clusters}, where we plot the generalized diffusion coefficient 
\begin{equation}
K_T(t) = \langle \mathbf{x}^2(t) \rangle_T  t^{-2/d_w},
\label{eq:Kt_T}
\end{equation}
as a function of time for 10 independent realizations of the clusters. We see a considerable scattering of these coefficients at any given time, and the absence of a clear trend or timescale. The corresponding mean (average over $N= 1000$ clusters) is shown by a dashed line tending to a horizontal asymptotic line. We note that the overall behavior of $K_T(t)$ (which is shown on log-linear scales) resembles log-periodic oscillations seen in many physical observables on regular fractals, with the important difference that now the oscillations are not periodic on a log-scale but random. 

\begin{figure}[h!]
\centering
\def\svgwidth{0.65\columnwidth}
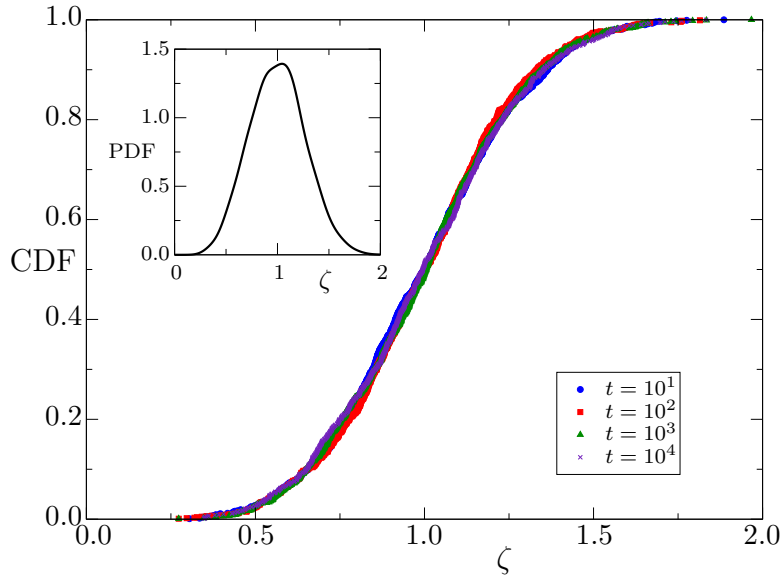
\caption{Empirical cumulative distribution function (CDF) of the random variable $\zeta(t)$ clearly showing universal fluctuations. The inset shows its probability density function (PDF) obtained by the kernel density approximation. }
\label{fig:CDF}
\end{figure}

The MSD (average over thermal histories) in an ensemble of clusters shows the universal fluctuation. To assess the corresponding distribution we take a closer look at the statistical properties of the generalized diffusion coefficient $K_T(t)$. To do so, we fix four different values of time, $t=10, 100, 1000$ and $10000$, and compute the MSD for $N = 1000$ different clusters. These MSDs are normalized with respect to the double-averaged MSD at the corresponding time giving a random variable
\begin{equation}
\zeta(t) = \frac{\langle \mathbf{x}^2(t) \rangle_T}{\langle \mathbf{x}^2(t) \rangle} = \frac{K_T(t)}{K(t)}.
\label{eq:scale}
\end{equation}
The empirical cumulative distribution function (CDF) of $\zeta(t)$ is shown in Fig.~\ref{fig:CDF}. The rescaled effective diffusion coefficients at different times indeed show a universal fluctuation: a Kolmogorov-Smirnov test \cite{Wasserman2004} performed at a significance level $\alpha =  0.05$ fails to reject the null hypothesis: All CDFs come from the same distribution. The inset of Fig.  \ref{fig:CDF} shows the kernel density approximation to the PDF of $\zeta$ for which we pooled all 4000 data points for 1000 clusters and four different time values (the kernel width $h= 1.26 \; \hat{\sigma} \; n^{-1/5}$ with $\hat{\sigma}$ the standard deviation of the sample  was chosen slightly larger than the recommended value of  $1.06 \; \hat{\sigma} \; n^{-1/5}$ \cite{Wasserman2006}, at which the PDF still showed perceptible fluctuations). This PDF gives us a flavour of the strength of fluctuations, i.e. of how different percolation clusters are with respect to diffusion thereon. Due to normalization, the mean of $\zeta$ is unity. Its variance is $0.072$ (which corresponds to the standard deviation of $\sigma = 0.27$), and its skewness has a positive value of $0.135$, meaning that the distribution is right-skewed, which is no wonder, since its PDF has to vanish for $\zeta < 0$. The distribution of $\zeta$ is platykurtic: its excess kurtosis is $-0.136$. 

\begin{figure}[h!]
\centering
\def\svgwidth{0.65\columnwidth}
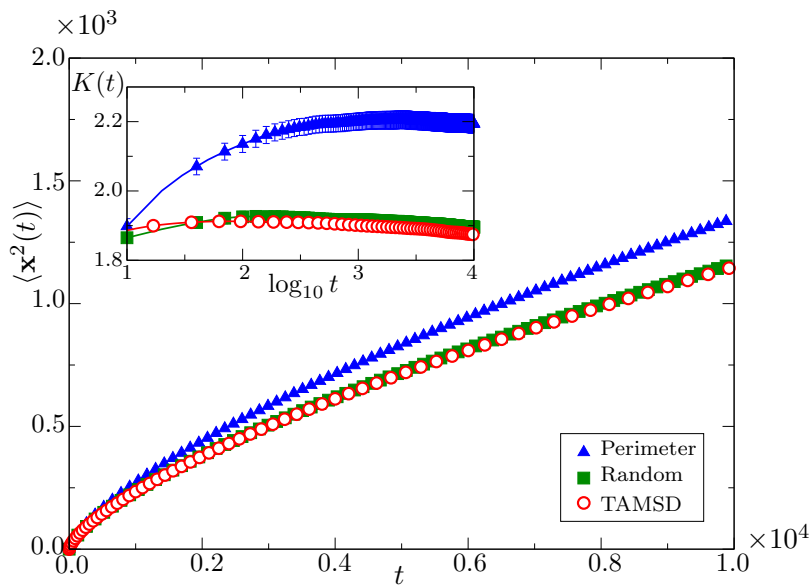
\caption{Mean Square Displacement (MSD) as a function of time computed in three different ways: \textit{triangles:} landscape average of the thermal history MSD when starting on the perimeter sites of the infinite cluster where considered, \textit{squares:} landscape average of the thermal history MSD with initial conditions taken at random within the inner square of lattice, and \textit{circles:} The time averaged MSD (TAMSD) averaged over the realizations of percolation clusters. The generalized diffusion coefficients for all the cases are shown in the inset.}
\label{fig:MSD}
\end{figure}

The fluctuations of $K_T(t)$ are not quite unpredictable, and depend on initial conditions, which are never forgotten in course of the temporal evolution. To see this we compare different procedures of finding the initial site, the diffusion seed, on a given realization of the infinite cluster. The procedure leading to the values of the double average $\langle \mathbf{x}^2(t) \rangle$ mentioned above is as follows. After generating a cluster on a sufficiently large lattice of a size $1000 \times 1000$ (see \ref{AppLand} for details), the initial site was chosen at random within a central inner square of size $300 \times 300$. If this site didn't belong to the cluster, another site was chosen at random, and the procedure was repeated until a site belonging to the cluster was found. If such site was not found, the whole configuration was discarded. Then a new cluster was generated and the whole process is repeated until a suitable initial site is found. The corresponding values of the MSD $\langle \mathbf{x}^2(t) \rangle$ are plotted in Fig. \ref{fig:MSD} with green squares. This one coincides with the time-averaged MSD (TAMSD, red circles) obtained over long trajectories, thus indicating the ergodicity of the process, property which
 will be discussed in  more detail in Sec.~\ref{Sec:Ergodic}.
 
Now let us consider the following, different procedure of finding the seed. The search for the seed starts from the center of the lattice. If the center of the landscape does belong to the infinite cluster, the central site is selected as the initial condition. On the other hand, if the center does not belong to the infinite cluster, then the closest point belonging to the cluster is iteratively searched for by considering all points in a square enclosing the center of the landscape (not only its first neighbors), then verifying one by one if they belong to the infinite cluster, and taking as the seed the first site found which belongs to the cluster. If none of these points belongs to the infinite cluster, a new square is considered that encloses the former one, and the process repeats until an initial site belonging to the cluster is found. The MSD obtained by this procedure  (not shown) differs considerably from the one obtained by the previous one and from the TAMSD. This finding urged us to consider the situation in more detail.

A more careful look at the results identifies the statistical reason for this. If the initial central site does belong to the infinite cluster, this site might be an internal or a perimeter site of the cluster.
If the central site does not belong to the infinite cluster, a site found by the search procedure is always a perimeter site. Applying the above searching procedure puts a larger statistical weight onto the perimeter sites than onto the internal ones since the probability that the center of the lattice is a cluster site is equal to the density of the infinite cluster, which is very small (asymptotically vanishing). Discarding all situations when the cluster site was found immediately, we may get the MSD for the situations, when the starting point is only on the perimeter. Discarding situations when we need for enlargement of a search area, i.e., those situation when the above searching algorithm should be applied, leads to the case when both perimeter and internal sites are weighted adequately. While the second situation reproduces the results obtained using the first method (random search in a central square), albeit with very poor statistics, the first one leads to a considerably larger MSD and thus larger effective diffusion coefficient. The MSD for the case when one starts at perimeter sites is shown in Fig.  \ref{fig:MSD} with blue triangles.

\section{Ergodic behavior \label{Sec:Ergodic}}
 
Apart from two different types of cluster averages corresponding to two types of initial conditions, Fig. \ref{fig:MSD} shows the moving-time average (red circles). This kind of MSD is obtained from the particle's positions $\mathbf{x}(t)$ sampled at $M$  evenly separated time instants $t_i = i \; \Delta t$ ($i = 0, \dots, M$) with $\Delta t = T/M$ and $T$ being the total simulation time for the trajectory:
\begin{equation}
\overline{\delta^2}(t_{lag} = m \Delta t ) = \frac{1}{ M - m } \sum_{i = 0}^{M-m} [ \mathbf{x}_{i + m} - \mathbf{x}_i ]^2.
\label{eq:TAMSD}
\end{equation}
The value of $\overline{\delta^2}(t_{lag} )$ for a finite trajectory is still a random variable, and  for long time lags, the statistical fluctuations in TAMSD around the double averaged MSD are large but show no systematic deviations, \textit{vide infra}. Fig.~\ref{fig:MSD} shows the results in which an additional averaging over an ensemble of clusters was performed to smoothen the curve. The results confirm the finding of \cite{Meroz2013} that the double average MSD and the TAMSD coincide within the statistical error. Therefore, strong fluctuations in diffusivity in different realizations (i.e. the absence of self-averaging, or, equivalently, of ergodicity with respect to the spatial organization of the system) coexist with the ergodicity of the overall process in the time domain. This ergodicity can essentially be anticipated on the grounds discussed in the Appendix B of Ref. \cite{Meroz2015}, and means that averaging over a sufficiently long trajectory is equivalent to averaging over the starting points of shorter ones.

The fact that in a single realization the initial position is never forgotten, however, strongly influences the art, how the convergence of the moving time average and ensemble double average takes place, reflecting itself in the temporal behavior of relative fluctuation of random TAMSD at finite data acquisition times (trajectory lengths) $T$. 

In the discussion of the ergodicity breaking or absence thereof, one often defines the ergodicity breaking parameter \cite{He2008,Deng2009}. To characterize convergence (or non-convergence) to the ergodic behavior, one considers the relative fluctuation of the TAMSD for a given time lag $t$ as a function of $T$,  i.e. its variance normalized on the square of its mean
\begin{equation}
EB(t,T) = \frac{ \left\langle \left( \overline{\delta^2} \right)^2 \right\rangle_C - \left\langle \overline{\delta^2} \right\rangle_C^2 }{ \left\langle \overline{\delta^2} \right\rangle_C^2 } = 
\frac{ \left\langle \left( \Delta \overline{\delta^2} \right)^2 \right\rangle_C}{ \left\langle \overline{\delta^2} \right\rangle_C^2 } 
\label{eq:EB}
\end{equation}
with $\Delta \overline{\delta^2} = \overline{\delta^2} - \left\langle \overline{\delta^2} \right\rangle_C$.

Ref.  \cite{Meroz2013} has demonstrated that the step-step autocorrelation function (ACF) in the percolation model with exponential waiting times on sites is practically indistinguishable from the corresponding ACF of increments of the fractional Brownian motion (FBM) with the corresponding Hurst exponent $H = 1/d_w < 1/2$. For FBM, this value of the Hurst exponent belongs to the domain, in which the relative fluctuation as a function of $t$ and $T$ behaves as $EB(t,T) \propto t/T$ for long $T$. This result of \cite{Deng2009} follows from a lengthy calculation, but can essentially be explained in very simple words, which will later be used to explain the crucial difference between the FBM and the diffusion on a percolation cluster. Our discussion first concentrates on FBM. 

Let us consider the value of the squared displacement during a time interval of duration $t$, starting at time $t'$: $\delta^2(t') = [x(t'+t)-x(t')]^2$. The correlation function $C_{\delta^2 \delta^2}(t',t'';t) = \left\langle [\delta^2(t') - \langle \delta^2 \rangle] [\delta^2(t'') - \langle \delta^2 \rangle] \right\rangle$ can then be derived from the displacement-displacement correlation function for FBM, being a Gaussian process, for which a four-point correlation function can be represented in terms of the two-point ones, which are known, and is related to the square of the correlation function of the underlying FBM. It reads $C_{\delta^2 \delta^2}(t',t'';t) = 2 K \{|t''-t'-t|^{2H} + |t''-t'+t|^{2H} -2|t''-t'|^{2H} \}^2$ with $K$ being the effective subdiffusion coefficient, and is only a function of a difference $\Delta t = t''-t'$ between the beginnings of the corresponding $t$-intervals during which the displacement is measured: $C_{\delta^2 \delta^2}(t',t'';t) = C_{\delta^2 \delta^2}(\Delta t;t)$. The normalized correlation function $\tilde{C}(\Delta t) =  C_{\delta^2 \delta^2}(\Delta t)/C_{\delta^2 \delta^2}(0)$ is bounded, non-negative, and decays for $\Delta t$ large as $\Delta t^{4H-4}$. This dimensionless function is essentially only a function of a dimensionless variable $\xi = \Delta t / t$. As long as it is integrable, one can define the typical correlation time $\tau = \int_0^\infty \tilde{C}(\Delta t) d \Delta t$ which is finite for $0< H <  3 / 4$ (and then proportional to $t$) and diverges otherwise. 
 
If the correlation time is finite, the behavior of the relative fluctuation of the random variable $\overline{\delta^2} $ is the same as the one of the sum of $N \sim T /\tau  \propto T /t $ independent random variables with the same variance, i.e. $EB(t,T) \propto 1 / N \sim t /T $. When the integrability of the ACF of displacements squared is lost, i.e. for $H \geq 4 /3$, one observes the change in this dependence. 

\begin{figure}[h!]
\centering
\def\svgwidth{0.65\columnwidth}
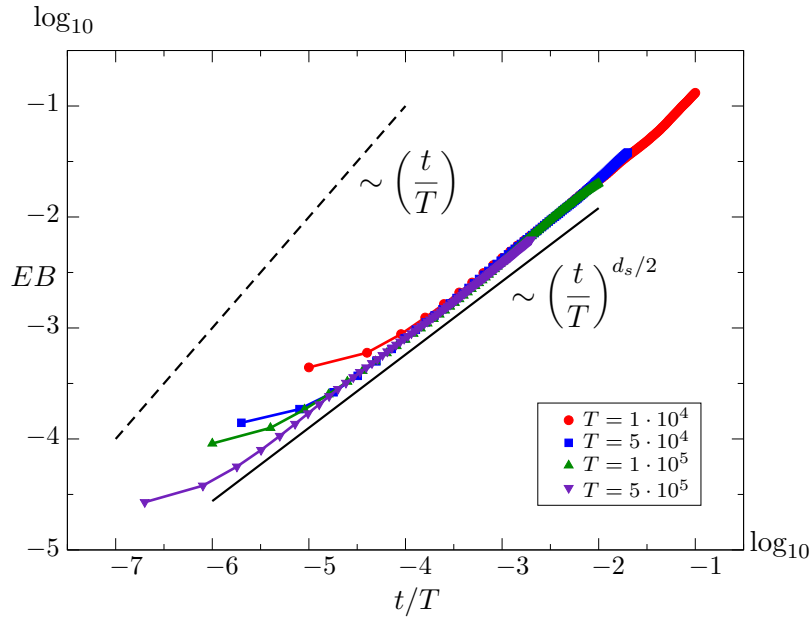
\caption{Time-dependent relative variance of the TAMSD (ergodicity breaking parameter) for three different trajectory's lengths $T$ as a function of the ratio between the lag time and the trajectory's length. 
The straight line gives its power law decay.}
\label{fig:EB}
\end{figure}

The behavior of the relative fluctuation $EB(t,T)$ for different values of $T$ and $t$ for the diffusion on the percolation cluster is shown in Fig. \ref{fig:EB}. The value of $EB$ is shown on double logarithmic scales as a function of $t / T$, as suggested by the previous discussion. As we see, the overall scaling with respect to $ t / T$ is well confirmed by numerical simulations. The deviations from the master curve visible at the lower parts of the corresponding curves corresponds to the values of $t$ of the order of typical waiting times, at which only few steps are made on the average. The slope of the master line is however considerably less than unity. Since this slope derives from the properties of the four-point correlation function of displacement, which for diffusion on a percolation cluster (evidently, not a Gaussian process even with respect to its single-point properties) may differ strongly from the one in FBM. 

This different slope of the master line is however easy to explain qualitatively in view of strong variations of the MSD of a random walker starting at different positions. This behavior corresponds closely to what is shown in Fig. \ref{fig:K_clusters}. When starting at some given point on a cluster, a walker explores during time $t$ on the average a part of a cluster with the number of sites $S$ depending on $t$ as $S \sim t^{d_s / 2}$ with $d_s$ being the spectral dimension of the incipient percolation cluster, which in two dimension is $d_s \approx 1.31$. Starting at a distant initial point, a walker explores a different domain, the one of a different local structure, and the displacements in such domains can be considered as independent. Now, it is known that a random walk on a fractal with $d_s < 2$ is recurrent, i.e. walkers revisit points on the fractal during a finite time. Therefore, when returning to the same domain (i.e. practically to the same starting point), the particle sees the same local environment, and has the squared displacement which is strongly correlated to the ones on previous visits. The MSD on multiple revisiting is different between different domains. Therefore one can assume that the total number of independent random variables to be averaged over when discussing the relative fluctuation is of the order of the number of such independent domains. Since the total number of different sites visited during time $T$ grows with $T$ as $S_{tot} \sim T^{d_s/2}$, the number of independent variables behaves as $N \approx S_{tot}/S \sim \left(T / t \right)^{ d_s / 2}$, and the relative fluctuation behaves as 
\begin{equation}
EB(t,T) \sim \left(\frac{t}{T} \right)^{\frac{d_s}{2}}. 
\label{eq:EB_asymp}
\end{equation}
The line with this slope drawn in Fig. \ref{fig:EB} describes the behavior observed very well. 
The possible practical importance of this finding is that it gives a very simple recipe to distinguish between the subdiffusion induced by intrinsic memory (like in FBM) and the one induced by disorder. The recipe is numerically simpler one than the ones discussed in \cite{Meroz2013} and \cite{Meroz2015}, does not imply additional coarse-graining, and could be preferable to other methods, if long enough trajectories are available. 

We note that including finite clusters into the analysis  \cite{Mardoukhi2015,Mardoukhi2018} change the results considerably. Such situations lead to an essentially non-ergodic behavior, and show richer dynamics due to the presence of finite clusters where particles can be trapped. However, the understanding of what happens on the infinite cluster alone is still important both because this situation is basic for the theoretical understanding of other cases, and because it can be realized experimentally. 

\section{Summary}

In this work, we revisited the problem of diffusion on the incipient infinite percolation cluster. Relying on results of extensive numerical simulations, we discuss the form of the PDF of displacements in such diffusion, and the way this PDF converges to its scaling form. This convergence takes place by narrowing of the central peak of the PDF, which pattern was recently proposed to be typical for systems with strong spatial disorder. The MSDs in single realizations of percolation clusters (or MSDs obtained when starting at distant initial positions on the same cluster) differ considerably. Moreover, these MSDs never forget the corresponding initial conditions, which can be seen e.g. from the fact, that even been averaged over the ensemble of clusters, the MSDs obtained when starting everywhere in the cluster or only on its perimeter sites differ significantly and systematically. The distributions of MSDs at fixed times show universal fluctuations. Although the absence of self-averaging of the MSD in percolation was reported long ago, the properties of the corresponding distribution, up to our best knowledge, were never reported on. This absence of self-averaging goes however hand in hand with the temporal ergodicity of the diffusion process. The art of decay of relative fluctuations of time-averaged MSD at a given lag time for growing data acquisition times differs from the one in similar processes, like fractional Brownian motion with the same Hurst exponent. This specific are of decay can be used as an additional test discerning between the subdiffusion induced by intrinsic memory (like in FBM) and the one induced by disorder. 

\section*{Acknowledgments}
The work of A.P.P. was financially supported by Doctoral  Programmes  in  Germany  funded  by  the Deutscher  Akademischer  Austauschdienst  (DAAD) (Programme ID 57440921).

\appendix

\section{Numerical methods}

The numerical simulations performed in this work were similar to those ones in \cite{Pacheco2021}. Here, they are explained for completeness.  

\subsection{Landscapes and averages \label{AppLand}}

A Bernoulli bond percolation system is constructed on a square lattice with a concentration of bonds equal to the critical concentration of $p_c= 0.5$. The process of generating the landscapes where the particles will diffuse, i.e., of the infinite clusters, consists of two stages: first, a large percolation system is generated by taking the bonds to be intact with probability $p = p_c = 0.5$, and broken otherwise. Once the percolation system is created, the Hoshen-Kopelman algorithm \cite{Hoshen1976} is  used  to  label  all the clusters present in the system. Then the largest (“infinite”) cluster is spotted out, and the rest erased. This infinite cluster is then the landscape for diffusion. The  sites not  belonging  to  this infinite  cluster  are declared unavailable, and the transition probabilities to them are set to zero.  The sizes of the systems used in this work were between $800 \times 800$ and $1000 \times 1000$ so that  particles will not hit the borders of the system during the simulation time. On  each  landscape  (i.e.   cluster),  $10000$  particles were simulated to generate the PDF and thermal history MSD in a single realization, starting from the same initial site, whose choices are described in the main text. For the moving time average only one particle was considered. Then, a  mean  over  $5000$  realizations  of  the  landscapes was taken to obtain the mean PDF shown in Fig. \ref{fig:PDF} of the main text. For the cluster average MSD, both for thermal histories and moving time averages, a mean over $1000$ landscapes was taken. The result of this can be seen in Fig. \ref{fig:MSD}. For the study of ergodicity breaking parameter, one particle was simulated per landscape, and a mean over $500$ landscapes was used to obtain Fig. \ref{fig:EB}.

\subsection{Trajectories \label{AppTraj}}

The diffusion of the particles is modeled as a continuous  time  random  walk  with  exponentially  distributed waiting time on the sites of a simulation lattice.  The trajectories  of  the  particles  diffusing  in  the  infinite percolation cluster  were  obtained  using  the  Gillespie  algorithm \cite{Toral2014}. Being on a site $\sigma= (i,j)$, the particle can jump to one of the four neighboring sites $\sigma \to \delta_i$, with $\delta_1= (i+ 1,j)$, $\delta_2= (i-1,j)$, $\delta_3= (i,j+ 1)$ and $\delta_4= (i,j-1)$, if available,  with  transition  rates  $\omega(\sigma \to \delta_i)$.  These transition rates from a site to each of $m$ accessible  neighboring  sites  are  taken  to  be  the  same and equal to unity, $\omega(\sigma \to \delta_i) = 1$, while the transition probabilities to forbidden sites are set to zero.  This corresponds to different jumping probabilities at different sites, $p(\sigma \to \delta_i) = 1/m$, and a total jumping rate $\omega(\sigma) = m$. Under these 
condition the equilibrium distribution (in any finite realization of the system) is homogeneous. 

A  particle  starts  at $t_0 = 0$  from  an initial  position $\sigma_0$.  At each simulation step $n$, corresponding to time $t_n$, the following procedure is repeated.  The jumping rate $\omega(\sigma_n)$ on the site $\sigma_n$ defines the exponentially distributed waiting time $\tau_n$,  after  which  the  next  jump  takes  place.   At a jump, the particle changes to one of the neighboring sites with the corresponding transition probability.  The time is increased to $t_{n+1} = t_n +\tau_n$.  The procedure is repeated until  the  maximum  simulation  time $t_{\max}$  is achieved.  One ends up with a list of particle’s positions and their respective jumping times.  This list is sampled to get particle’s positions at time intervals $\Delta t= 0.1$ which then is used for calculating all the needed quantities. The value of $t_{\max}$ depends on the simulation: for the PDF, a $t_{\max} = 12000$, for the thermal history MSD, $t_{\max} = 10000$ and for the TAMSD $t_{\max} = 50000$ with a maximal time lag $t_{lag}^{(\max)} = 10240$.

\section*{References}

\bibliography{JPA_percolation}

\end{document}